\documentclass[prb,aps,superscriptaddress,showpacs,twocolumn,amssymb,amsmath]{revtex4}
\usepackage{graphicx}
\usepackage{theorem}
\usepackage{dcolumn}
\usepackage{bm}
\usepackage{mathrsfs}
\usepackage{amsmath}    
\usepackage{setspace}

\begin{document}

\title{Shape fluctuations and optical transition of He$_{2}^{*}$ excimer tracers in superfluid $^4$He}
\author{W. Guo\footnote{Corresponding: wguo@magnet.fsu.edu}}
\affiliation{National High Magnetic Field Laboratory, 1800 East Paul Dirac Drive, Tallahassee, FL 32310, USA}
\affiliation{Mechanical Engineering Department, Florida State University, Tallahassee, FL 32310, USA}

\author{A. I. Golov}
\affiliation{School of Physics and Astronomy, The University of Manchester, Manchester M13 9PL, United Kingdom}

\date{\today}

\begin{abstract}
Metastable He$_{2}^{*}$ excimer molecules have been utilized as tracer particles of the normal component in superfluid $^4$He (He II) which can be imaged via laser-induced fluorescence. These excimer molecules form tiny bubbles in He II and can bind to quantized vortices at sufficiently low temperatures, thereby allowing for direct visualization of vortex dynamics in an inviscid superfluid. However, the $a^{3}\Sigma^+_{u}$${\rightarrow}$$c^{3}\Sigma^+_{g}$ optical absorption line, which is responsible for the fluorescence imaging of the He$_{2}^{*}$ molecules, is controlled by fluctuations on the bubble shape, and its exact line profile is not known at low temperatures. In this paper, we present a bubble model for evaluating the surface fluctuation eigenmodes of the excimers in He II. The line profile of the $a^{3}\Sigma^+_{u}{\rightarrow}c^{3}\Sigma^+_{g}$ transition is calculated at different temperatures by considering both the zero-point and thermal fluctuations on the bubble shape. We show that, as the temperature drops from 2~K to 20 mK, the peak absorption strength is enhanced by a factor of about five, accompanying a blueshift of the peak location by about 2 nm. A double-peak line profile due to the rotational levels of the molecular core can be resolved. This bubble model also allows us to evaluate the stiffness of the He$_{2}^{*}$ bubbles and hence their diffusion constant in He II due to scattering off thermal phonons. Our results will aid the design of future experiments on imaging quantized vortices in He II using He$_{2}^{*}$ tracers.
\end{abstract}

\pacs{67.25.dk, 29.40.Gx, 47.27.-i} \maketitle
\section{Introduction}\label{Sec1}
The dynamics of quantized vortex lines in a coherent matter-wave system is responsible for a wide range of phenomena, such as the decay of quantum turbulence \cite{Vinen-2006-JLTP, Nemirovskii-2013-PR} and the initiation of dissipation in type-II superconductors \cite{Anderson-1964-RMP, Larbalestier-2001-Nature}, and is also implicated in the appearance of glitches in neutron star rotation \cite{Anderson-1975-Nature, Packard-1972-PRL} and the formation of cosmic strings in the early universe \cite{Zurek-1985-Nature}. A systematic study of vortex-line dynamics promises broad significance spanning multiple physical science disciplines. A powerful method to study vortex-line motion is via direct line visualization in a superfluid, which can be achieved in both superfluid helium and atomic Bose-Einstein condensates (BECs) \cite{Tsubota-2013-PR}. However, given their small sample sizes (typically $\sim$10$^2$ $\mu$m in diameter), experimental studies of vortex-line dynamics and quantum turbulence in BECs are only just emerging \cite{Henn-2009-PRL, Navon-2016-Nature} with some active research focusing on two-dimensional quantum turbulence \cite{Johnstone-2019-Science, Gauthier-2019-Science}. In contrast, superfluid helium at low temperatures provides an ideal system for laboratory studies of quantum turbulence that spans many orders of magnitude in length scale.

However, experimental observation of vortex lines in superfluid helium is very challenging due to the angstrom-sized cores of the vortices \cite{Donnelly-1991-book}. Instead of imaging the thin vortex lines directly, a number of efforts have been devoted to decorating quantized vortex lines with tracer particles for line visualization. For instance, Yarmchuk \emph{et al.} imaged the termination of rectilinear vortex lines on the free surface of superfluid $^4$He (He II) by photographing the electrons pulled along the vortex lines to the free surface \cite{Yarmchuk-1979-PRL}. Guo \emph{et al.} reported a method to image electron bubbles trapped on vortices in bulk He II via acoustic cavitation of these bubbles \cite{Guo-2007-JLTP-bubble,Guo-2009-PRB}. However, the heating due to the transducer vibration and the strong acoustic waves can severely disturb the fluid. More recently, Bewley \emph{et al.} used micron-sized frozen hydrogen particles to decorate vortices and successfully visualized vortex lines in He II \cite{Bewley-2006-Nature}. This team has since filmed real-time vortex-line reconnections and Kelvin waves on vortices \cite{Bewley-2008-PNAS,Paoletti-2008-PRL,Fonda-2014-PNAS,Fonda-2019-PNAS}. Nevertheless, the injection of the hydrogen particles is usually accompanied by a large heat load which limits the application of this technique to above about 1.6 K where the vortex dynamics can be strongly affected by the viscous normal-fluid component in He II. On the other hand, there is an evolving interest in imaging quantized vortices at lower temperatures in a pure superfluid. For instance, a key question in quantum turbulence research is how the energy of a vortex tangle decays in a pure superfluid with zero viscosity \cite{Vinen-2006-JLTP}.

So far, there have been two efforts in imaging quantized vortices in superfluid helium with minimal normal-fluid fraction. One method is to image quantum turbulence in superfluid $^3$He-B via Andreev reflection of quasiparticles \cite{Ahlstrom-2014-JLTP, Baggaley-2015-PRL}. This method is still at an early stage of development. The other method, adopted by Gomez \emph{et al.}, is to dope a beam of fast-moving $^4$He nanodroplets with xenon atoms \cite{Gomez-2014-Sci}. These droplets are evaporatively cooled to 0.38 K and the vortices in them can be imaged via X-ray diffraction of the trapped xenon atoms. Nevertheless, this experiment can only generate one-time snapshots of the vortices and does not allow dynamical study of the vortex motion.

On the other hand, the feasibility of using He$^*_2$ excimer molecules as tracers in He II has been validated through a series of experiments \cite{Guo-2009-PRL,Guo-2010-JLTP,Guo-2010-PRL}. These molecules can be created easily as a consequence of ionization or excitation of ground state helium atoms \cite{Benderskii-2002-JCP} and can be imaged via a laser-induced fluorescence (LIF) technique \cite{Guo-2009-PRL,McKinsey-2005-PRL,Rellergert-2008-PRL}. These excimers in the electron-spin triplet ground state $a^3\Sigma^{+}_u$ have an exceptional 13-second radiative lifetime \cite{McKinsey-1999-PRA}, and they form tiny bubbles in liquid helium (about 6 {\AA} in radius \cite{Benderskii-2002-JCP}). Due to their small size and hence small binding energy to the vortex cores \cite{Mateo-2015-JCP}, above 1 K, He$^*_2$ molecules are solely entrained by the viscous normal fluid in He II, which allows for quantitative study of the normal-fluid velocity field \cite{Marakov-2015-PRB,Gao-2016-JETP,Gao-2016-PRB,Gao-2017-PRB,Gao-2018-PRB}. Furthermore, it has been demonstrated by Zmeev \emph{et al.} that below about 0.2 K, the He$^*_2$ tracers can permanently bind to quantized vortices \cite{Zmeev-2013-PRL}, thereby enabling vortex-line imaging in the absence of the normal fluid in He II.

The fluorescence imaging of the He$^*_2$ molecules is essentially controlled by the $a^{3}\Sigma^{+}_{u}$${\rightarrow}$$c^{3}\Sigma^{+}_{g}$ optical absorption transition (see discussions in Sec.~\ref{Sec2-1}). The peak wavelength and the strength of this transition can be affected by fluctuations on the He$^*_2$ bubble shape. As the temperature drops, the reduced bubble surface fluctuations may lead to a sharper transition line profile with a slightly shifted peak wavelength. However, despite extensive measurements of the He$^*_2$ optical transitions at relatively high temperatures \cite{Hill-1971-PRL,Eltsov-1998-JLTP,Kafanov-2000-JETP}, there is no data of the $a^{3}\Sigma^+_{u}$${\rightarrow}$$c^{3}\Sigma^+_{g}$ absorption line below 1 K. To guide future vortex-line imaging experiments using He$^*_2$ tracers, we hereby present a theoretical study of the He$^*_2$ bubble surface fluctuations. In Sec.~\ref{Sec2}, we discuss the optical transitions for fluorescence imaging of the He$^*_2$ molecules, the bubble model, and the eigenmodes of the bubble surface fluctuations. In Sec.~\ref{Sec3}, we discuss how the $a^{3}\Sigma^{+}_{u}$${\rightarrow}$$c^{3}\Sigma^{+}_{g}$ absorption line varies as the temperature drops from 2~K to 20 mK. In Sec.~\ref{Sec4}, we evaluate the diffusion of the He$^*_2$ molecules in He II using the calculated bubble stiffness and compare the result with experimental observations. A brief summary is given in Section~\ref{Sec5}.

\section{Modeling He$^*_2$ molecular bubbles in He II}\label{Sec2}
\subsection{Fluorescence imaging of He$^*_2$ molecules}\label{Sec2-1}
\begin{figure}[htb]
\includegraphics[scale=0.5]{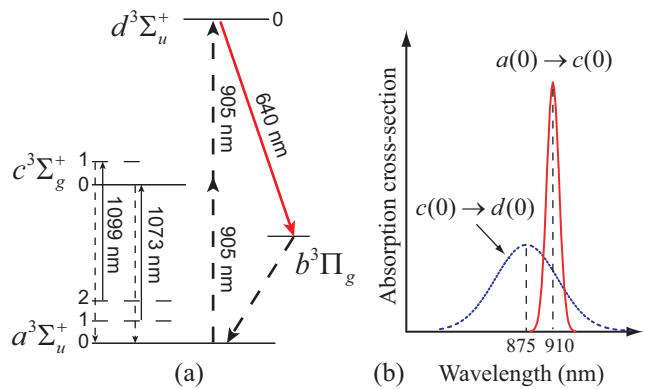}
\caption{(color online). (a) Schematic diagram showing the cycling transitions for imaging the He$^*_2$ triplet molecules. The levels labeled 0, 1, 2 are the vibrational levels for each corresponding electronic state; (b) Schematic diagram showing the $a$(0)${\rightarrow}$$c$(0) and $c$(0)${\rightarrow}$$d$(0) absorption transition lines.} \label{Fig1}
\end{figure}

In order to image the He$^*_2$ molecules in the $a^{3}\Sigma^{+}_{u}$ triplet ground state, McKinsey's group first developed a cycling-transition LIF technique \cite{McKinsey-2005-PRL,Rellergert-2008-PRL,Rellergert-Dissertation}. The concept of this technique is shown schematically in Fig.~\ref{Fig1} (a). An infrared pulsed laser at 910$\pm$6 nm (with the maximum efficiency of the two-step transition at 905 nm) can be used to illuminate the He$^*_2$($a^{3}\Sigma^{+}_{u}$) molecules in He II. A molecule absorbs a 905-nm photon will undergo a transition from the zeroth vibrational level $a$(0) of the $a^{3}\Sigma^{+}_{u}$ state to the corresponding vibrational level $c$(0) of the $c^{3}\Sigma^{+}_{g}$ state. When the photon flux is sufficiently high, this molecule can subsequently absorb another 905-nm photon and transit to the d$^{3}\Sigma_{u}^{+}$ state before it radiatively decays back to the $a$(0) level. Calculations of the branching ratios indicate that only about 10\% of the excited molecules in the d$^{3}\Sigma_{u}^{+}$ state decay to the c$^{3}\Sigma_{g}^{+}$ state \cite{Rellergert-Dissertation}, while the remaining 90\% decay to the b$^{3}\Pi_{g}$ state, emitting detectable red photons at 640 nm. Molecules in both the c$^{3}\Sigma_{g}^{+}$ and b$^{3}\Pi_{g}$ states then decay back to the a$^{3}\Sigma_{u}^{+}$ state, and the process can be repeated. Since the emitted photons are well separated in wavelength from the excitation photons, scattered 905-nm pump laser light can be blocked by appropriate filters. However, in the cycling transitions the molecules may fall into the long-lived $a$(1) and $a$(2) vibrational levels of the $a^{3}\Sigma^{+}_{u}$ state, where they are off-resonant to the 905-nm pump laser and are lost for subsequent cycles \cite{Rellergert-2008-PRL, Rellergert-Dissertation}. To recover the lost molecules, two continuous lasers at 1073~nm and 1099~nm are normally used to repump the molecules in the $a$(1) and $a$(2) vibrational levels to the $c$(0) and $c$(1) states, respectively. The molecules in these states rapidly decay back to the triplet ground $a$(0) state and can be reused.

At first sight, this cycling transition scheme could be improved by replacing the single 905-nm pump laser by two pulsed lasers at the peak resonance wavelengths of the $a$(0)${\rightarrow}$$c$(0) and the $c$(0)${\rightarrow}$$d$(0) transitions, respectively. Indeed, Rellergert has done systematic measurements of the optical absorption transitions of the He$^*_2$($a^{3}\Sigma^{+}_{u}$) molecules in He II \cite{Rellergert-Dissertation}. It turns out that, as shown in the schematic in Fig.~\ref{Fig1} (b), the $a$(0)${\rightarrow}$$c$(0) absorption line is centered at about 910 nm and has a narrow line profile with a full width at half maximum (FWHM) of about 20 nm in He II around 2 K. On the other hand, the $c$(0)${\rightarrow}$$d$(0) transition, centered around 875 nm in He II, has a much broader profile with a FWHM of the order 100 nm. In the cycling transition of the He$^*_2$ molecules, this $c$(0)${\rightarrow}$$d$(0) absorption line could be even broader due to the shape relaxation of the molecular bubbles toward the equilibrium $c$(0) shape following the $a$(0)${\rightarrow}$$c$(0) transition. Therefore, the absorption cross-section of the $c$(0)${\rightarrow}$$d$(0) transition does not decrease much as the excitation wavelength changes from the peak wavelength at 875 nm to 905 nm. Practically, the gain of the two-color excitation scheme is marginal due to the timing jitter and imperfect spatial overlap of the two laser pulses. As Rellergert concluded, a single pump laser at 905 nm is the optimal choice since a single beam is always perfectly overlapped with itself in both space and time.

Through the above discussions, it is clear that the LIF imaging of the He$^*_2$($a^{3}\Sigma^{+}_{u}$) molecules is dominantly controlled by the $a$(0)${\rightarrow}$$c$(0) absorption transition. A subsequent question is how this transition line profile varies as the He II is cooled to far below 1 K. Obviously, the width and the peak wavelength of this absorption line can be affected by He$^*_2$ bubble shape fluctuations. To study this effect, we will adopt a He$^*_2$ bubble model.

\subsection{Bubble model}\label{Sec2-2}
The strong repulsion between the Rydberg electron of a He$^*_2$ excimer molecule and the closed-shell $^4$He atoms can lead to the formation of a small bubble surrounding the He$^*_2$ molecule. The detailed structures of such He$^*_2$ bubble states in liquid helium have been examined by Eloranta \emph{et al.} \cite{Eloranta-2001-JCP, Eloranta-2002-JCP} and Bonifaci \emph{et al.} \cite{Bonifaci-2016-JPCA} using sophisticated density functional calculations. However, this density functional framework is not convenient for studying the bubble surface fluctuation modes and their effects on the optical transitions at finite temperatures. On the other hand, a classic bubble model has been successfully applied to explain the motion and the absorption lines of electron bubbles \cite{Fowler-1968-PR, Lerner-1993-JLTP, Maris-2004-JLTP}, excited helium atoms \cite{Hickman-1975-PRB}, and other atomic bubbles in liquid helium \cite{Kanorsky-1994-PRB}. Eloranta \emph{et al.} \cite{Eloranta-2002-JCP} and Kafanov \emph{et al.} \cite{Kafanov-2000-JETP} also applied the bubble model to study the optical transitions of He$^*_2$ molecules, although Kafanov \emph{et al.} used an approximated interaction potential between the He$^*_2$($a^{3}\Sigma^{+}_{u}$) molecule and the ground state He atoms. Nevertheless, there was no study of the surface fluctuation modes of the He$^*_2$ molecules. In what follows, we will adopt the bubble model while incorporating some key information about the He$^*_2$-He interactions derived in the density functional work \cite{Eloranta-2001-JCP, Eloranta-2002-JCP, Bonifaci-2016-JPCA}.

In the bubble model, the liquid helium is treated as a continuous medium whose number density around the He$^*_2$($a^{3}\Sigma^{+}_{u}$) molecule can be described by the Jortner's trial function \cite{Jortner-1965-JCP,Hirioke-1965-JCP}:
\begin{equation}
\resizebox{.89\hsize}{!}{$\rho(\vec{r})=
\begin{cases}
0 & r\leq R_0\\
\rho_0\{1-[1+\alpha (r-R_0)]e^{-\alpha (r-R_0)}\} & r>R_0
\end{cases}$},
\label{Eq:Density}
\end{equation}
where $\rho_0$ is the number density far from the molecule bubble, and $\alpha$ and $R_0$ are tuning parameters that can be adjusted to minimize the bubble energy. For a sharp interface, $\alpha R_0\gg$1. One can introduce an effective bubble radius $R_b$ at the barycenter of the interface where the helium density varies from zero to its bulk value \cite{Hickman-1975-PRB,Kanorsky-1994-PRB}:
\begin{equation}
\int^{R_b}_{0}\rho(\vec{r})r^2dr=\int^{\infty}_{R_b}[\rho_0-\rho(\vec{r})]r^2dr.
\label{Eq:Radius}
\end{equation}
Combining Eq.~\ref{Eq:Density} and Eq.~\ref{Eq:Radius}, one can derive that:
\begin{equation}
R_b=R_0\left(1+\frac{6}{\alpha R_0}+\frac{18}{\alpha^2R^2_0}+\frac{24}{\alpha^3 R^3_0}\right)^{1/3}.
\label{Eq:Radius1}
\end{equation}
The total energy of the He$^*_2$ bubble is then given by \cite{Eloranta-2002-JCP}:
\begin{equation}
E=E_e+PV+\sigma S+\frac{\hbar^2}{8M_{He}}\int\frac{(\nabla\rho)^2}{\rho}d^3r,
\label{Eq:Energy}
\end{equation}
where $E_e$ is the energy due to the He$^*_2$ molecule inside the bubble, $P$ is the pressure in the liquid, $\sigma$ is the helium surface tension coefficient, $V$ and $S$ are the volume and the surface area of the bubble, respectively. For a spherical bubble, $V$=$4{\pi}R^3_b/3$ and $S$=$4{\pi}R^2_b$. The last term in Eq.~(\ref{Eq:Energy}) accounts for the interfacial quantum kinetic energy \cite{Eloranta-2002-JCP, Hickman-1975-PRB, Kafanov-2000-JETP}, where $\hbar$ is Planck's constant and $M_{He}$ is the mass of a $^4$He atom. This term is negligible for relatively large bubbles such as electron bubbles (i.e., about 2 nm in radius) but is appreciable for small bubbles such as the He$^*_2$ excimer bubbles.

\begin{figure}[!tb]
\includegraphics[scale=0.58]{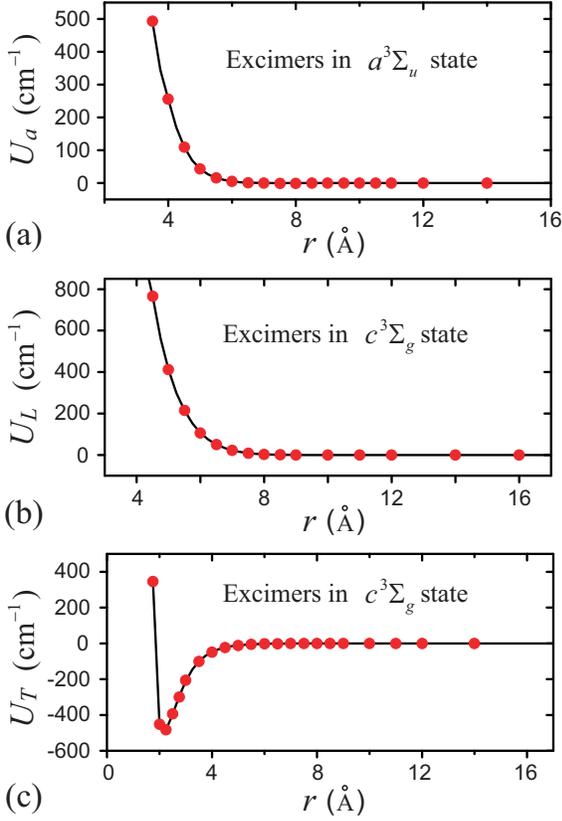}
\caption{Interaction potential between a He$^*_2$ excimer molecule and a ground state helium atom for the molecule (a) in the $a^{3}\Sigma^{+}_{u}$ state; (b) in the $c^{3}\Sigma^{+}_{g}$ state along the collinear direction of the molecular core; and (c) in the $c^{3}\Sigma^{+}_{g}$ state along the ``T'' direction as defined in ref.~\cite{Eloranta-2001-JCP}. Red dots are the data listed in Table II of ref.~\cite{Eloranta-2001-JCP}. Solid curves are our fits.} \label{Fig2}
\end{figure}

The molecule energy $E_e$ can be evaluated as the bare molecule energy in vacuum $E^{(vac)}_e$ plus the additional energy $E_{int}$ due to the interaction between the molecule and the surrounding helium when $E_{int}\ll E^{(vac)}_e$ \cite{Eloranta-2002-JCP,Kafanov-2000-JETP}:
\begin{equation}
E_e=E^{(vac)}_e+E_{int}=E^{(vac)}_e+\int{U_{int}(\vec{r})\rho(\vec{r})d^3r},
\end{equation}
where $U_{int}(\vec{r})$ is the He$^*_2$-He interaction potential. This $U_{int}(\vec{r})$ for various He$^*_2$ Rydberg states have been calculated by Eloranta and Apkarian using density functional method \cite{Eloranta-2001-JCP}. The equilibrium shapes and energies of these Rydberg states in liquid helium have also been derived and reported \cite{Eloranta-2002-JCP}. In order to incorporate these useful density functional results in our later analysis of the bubble surface modes, we extract the potential $U_{a}(\vec{r})$ between the He$^*_2$($a^{3}\Sigma^{+}_{u}$) molecule and a $^4$He atom by performing a least-squares fit to the discrete data listed in Ref.~\cite{Eloranta-2001-JCP} using simple polynomial functions of the form $U_{a}(\vec{r})$=$\sum^{n=12}_{n=1}C_nr^{-n}$, where the $C_n$'s are fitting parameters. $U_{a}(\vec{r})$ is treated as isotropic considering the nearly spherical symmetry of the Rydberg electron wavefunction in the molecule He$^*_2$($a^{3}\Sigma^{+}_{u}$). The result is shown in Fig.~\ref{Fig2} (a). We have also performed similar fits to the anisotropic interaction potential $U_c(\vec{r})$ between a He$^*_2$($c^{3}\Sigma^{+}_{g}$) molecule and a $^4$He atom since this information is needed in the evaluation of the $a\rightarrow c$ absorption line profile. Note that $U_c(\vec{r})$ can be expressed as~\cite{Eloranta-2002-JCP}:
\begin{equation}
U_c(\vec{r})=U_L(r)\cos^2(\theta)+U_T(r)\sin^2(\theta),
\end{equation}
where $\theta$ is the angle from the collinear direction of the molecular core. The fits to $U_L(r)$ and $U_T(r)$ are also shown in Fig.~\ref{Fig2}. As a convention in spectroscopy, energy is expressed in term of cm$^{-1}$. A conversion to real energy can be made by multiplying the value in cm$^{-1}$ by $hc$ (i.e., Planck's constant times the speed of light).

Knowing the interaction potential $U_{a}(\vec{r})$, one can then vary the tuning parameters $\alpha$ and $R_0$ to minimize the total energy $E$ and determine the equilibrium size of the He$^*_2$($a^{3}\Sigma^{+}_{u}$) bubble in liquid helium. For instance, at zero pressure, if we set $\rho_0$=0.02184 atoms/\AA$^3$ and $\sigma$=0.18 cm$^{-1}$/\AA$^2$ as used in Ref.~\cite{Eloranta-2002-JCP}, we get the equilibrium shape parameters $\alpha_{eq}$=1.8 \AA$^{-1}$ and $R_{0,eq}$=5.4 \AA, which corresponds to an equilibrium bubble radius $R_{b,eq}$=6.6 \AA.

\subsection{Surface fluctuation eigenmodes}\label{Sec2-3}
To study shape fluctuations of the He$^*_2$($a^{3}\Sigma^{+}_{u}$) bubble, we describe the deformed bubble shape as:
\begin{equation}
R_0(\theta,\phi)=R_{0,eq}\big(1+\sum_{l,m}\epsilon_{lm}Y_{lm}(\theta,\phi)\big),
\label{Eq:deformation}
\end{equation}
where $Y_{lm}(\theta,\phi)$'s are spherical harmonics, and $\epsilon_{lm}$'s are complex numbers that denote the amplitudes of the deformation modes. $\epsilon_{lm}$ satisfies $|\epsilon_{lm}|\ll$1 and $\epsilon_{l,-m}$=$-\epsilon^*_{lm}$ so that $R_0(\theta,\phi)$ is always a real number. In general, an angle-dependent shape parameter $\alpha(\theta,\phi)$ is also expected in the Jortner's density profile. However, significant variations of the interfacial thickness (and hence $\alpha$) can take place only when the shape deformation occurs at length scales comparable to $\alpha^{-1}_{eq}$. For a deformation mode $Y_{lm}(\theta,\phi)$, the length scale of the deformation is $2\pi R_{0,eq}/l$. We will see in later discussions that the stiffness of a deformation mode increases as $l^2$. Therefore, the mean amplitudes of the modes with large $l$ (and hence small deformation scales) are negligible in the examined temperature range. As a result, it is reasonable to neglect the angle variation of $\alpha$ and just take its equilibrium value $\alpha_{eq}$ in the subsequent analysis.

For the deformed bubble described by Eq.~(\ref{Eq:deformation}), the effective radius $R_b(\theta,\phi)$ now depends on the solid angle. Keeping the terms in Eq.~(\ref{Eq:Radius1}) to the second order in $\epsilon_{lm}$, we can write $R_b(\theta,\phi)$ as:
\begin{equation}
\resizebox{1\hsize}{!}{$
\begin{split}
&R_b(\theta,\phi)=R_{b,eq}\\
&~+\frac{R^3_{0,eq}}{R^2_{b,eq}}\bigg[1+\frac{4}{\alpha_{eq}R_{0,eq}}+\frac{6}{\alpha^2_{eq}R^2_{0,eq}}\bigg]\big(\sum_{l,m}\epsilon_{lm}Y_{lm}\big)\\
&~+\frac{R^6_{0,eq}}{R^5_{b,eq}}\bigg[\frac{2}{\alpha^2_{eq}R^2_{0,eq}}+\frac{12}{\alpha^3_{eq}R^3_{0,eq}}+\frac{12}{\alpha^4_{eq}R^4_{0,eq}}\bigg]\big(\sum_{l,m}\epsilon_{lm}Y_{lm}\big)^2.
\label{Eq:Rb-purt}
\end{split}
$}
\end{equation}

Due to the shape deformation, the total energy of the bubble would increase and to the lowest order in $\epsilon_{lm}$ can be expressed as:
\begin{equation}
\Delta E=E-E_{eq}=\sum_{l,m}\frac{1}{2}k_{lm}|\epsilon_{lm}|^2,
\label{Eq:dE}
\end{equation}
where $k_{lm}$ denotes the mode stiffness. Note that Eq.~(\ref{Eq:dE}) should not contain any first order terms in $\epsilon_{lm}$ since we consider bubble deformation around its equilibrium minimum-energy shape. To determine $k_{lm}$, we now evaluate the energy terms in Eq.~(\ref{Eq:Energy}).

To the second order in $\epsilon_{lm}$, the surface and volume energy terms are given by~\cite{Rayleigh-1945-book}:
\begin{equation}
\resizebox{1\hsize}{!}{$
\begin{split}
&\sigma{\int}dS=\sigma{\int}d\Omega\cdot R_b\sqrt{R^2_b+(\partial{R_b}/\partial\theta)^2+(\partial{R_b}/\partial\phi)^2/\sin^2\theta}\\
&~=4{\pi}R^2_{b,eq}\sigma+2\sqrt{4\pi}\sigma\frac{R^3_{0,eq}}{R_{b,eq}}
\bigg[1+\frac{4}{\alpha R_{0,eq}}+\frac{6}{\alpha^2R^2_{0,eq}}\bigg]\epsilon_{00}\\
&~+\sigma\frac{R^6_{0,eq}}{R^4_{b,eq}}\sum_{l,m}\bigg[\frac{l^2+l+2}{2}\bigg(1+\frac{4}{\alpha_{eq}R_{0,eq}}+\frac{6}{\alpha^2_{eq}R^2_{0,eq}}\bigg)^2\\
&~+\bigg(\frac{4}{\alpha^2_{eq}R^2_{0,eq}}+\frac{24}{\alpha^3_{eq}R^3_{0,eq}}+\frac{24}{\alpha^4_{eq}R^4_{0,eq}}\bigg)\bigg]\cdot|\epsilon_{lm}|^2,
\label{Eq:E_S}
\end{split}
$}
\end{equation}
\begin{equation}
\resizebox{1\hsize}{!}{$
\begin{split}
&P{\int}dV=P{\int}d\Omega\frac{1}{3}R^3_b(\theta,\phi)\\
&=\frac{4\pi}{3}R^3_{b,eq}P+\sqrt{4\pi}PR^3_{0,eq}\bigg(1+\frac{4}{\alpha_{eq}R_{0,eq}}+\frac{6}{\alpha^2_{eq}R^2_{0,eq}}\bigg)\epsilon_{00}\\
&+PR^3_{0,eq}\sum_{l,m}\bigg(1+\frac{2}{\alpha_{eq}R_{0,eq}}\bigg)\cdot|\epsilon_{lm}|^2.
\label{Eq:E_V}
\end{split}
$}
\end{equation}
The interfacial quantum kinetic energy term can be integrated to give:
\begin{equation}
\resizebox{0.88\hsize}{!}{$
\begin{split}
E_I&=E_{I,eq}+\frac{\hbar^2\rho_0R_{0,eq}}{8M_{He}}(7.28+10.31\alpha_{eq}R_{0,eq})\epsilon_{00}\\
&+\frac{1.45\hbar^2\rho_0R_{0,eq}}{8M_{He}}\alpha_{eq}R_{0,eq}\sum_{l,m}\frac{l^2+l+1}{2}\cdot|\epsilon_{lm}|^2.
\label{Eq:E_I}
\end{split}
$}
\end{equation}

As for the energy $E^{(a)}_e$ of the He$^*_2$($a^{3}\Sigma^{+}_{u}$) molecule in liquid helium, there is no analytic formula for evaluating its change when the bubble shape deforms. Nevertheless, considering the spherical symmetry of the interaction potential $U_a(\vec{r})$, we can write this energy term as:
\begin{equation}
E^{(a)}_e=E^{(a)}_{e,eq}+k^{(a)}_0\epsilon_{00}+\sum_{l,m}\frac{1}{2}k^{(a)}_{lm}|\epsilon_{lm}|^2.
\label{Eq:E_m}
\end{equation}
The coefficients $k^{(a)}_0$ and $k^{(a)}_{lm}$ can be determined numerically. For example, one may set the shape of the molecule bubble to be $R_0(\theta,\phi)$=$R_{0,eq}$[1+$\epsilon_{00}Y_{00}(\theta,\phi)$]. The integral ${\int}U_a(\vec{r})\rho(\vec{r})d^3r$ can be calculated numerically as a function of $\epsilon_{00}$, and the result can then be fitted with the function $E^{(a)}_{e,eq}+k^{(a)}_0\epsilon_{00}+\frac{1}{2}k^{(a)}_{00}|\epsilon_{00}|^2$ to determine the coefficients $k^{(a)}_0$ and $k^{(a)}_{00}$. We also perform similar fits to determine the first-order expansion coefficients for the $c^{3}\Sigma^{+}_{g}$ state since these coefficients are needed in later absorption line calculations:
\begin{equation}
E^{(c)}_e=E^{(c)}_{e,eq}+k^{(c)}_0\epsilon_{00}+k^{(c)}_2\epsilon_{20}.
\label{Eq:E_m_c}
\end{equation}
Note that due to the symmetry of the potential $U_c(\vec{r})$, there are two first-order terms in Eq. (\ref{Eq:E_m_c}). Typical values of these coefficients are listed in Table~\ref{Table-1}.


Collecting the results shown in Eq.~(\ref{Eq:E_S})-(\ref{Eq:E_m}), the stiffness $k_{lm}$ of each surface fluctuation mode can be determined:
\begin{equation}
\resizebox{1.0\hsize}{!}{$
\begin{split}
&k_{lm}=k^{(a)}_{lm}+2\sigma\frac{R^6_{0,eq}}{R^4_{b,eq}}\bigg[\frac{l^2+l+2}{2}\bigg(1+\frac{4}{\alpha_{eq}R_{0,eq}}+\frac{6}{\alpha^2_{eq}R^2_{0,eq}}\bigg)^2\\
&+\bigg(\frac{4}{\alpha^2_{eq}R^2_{0,eq}}+\frac{24}{\alpha^3_{eq}R^3_{0,eq}}+\frac{24}{\alpha^4_{eq}R^4_{eq}}\bigg)\bigg]+2PR^3_{0,eq}\bigg(1+\frac{2}{\alpha_{eq}R_{0,eq}}\bigg)\\
&+\frac{1.45\hbar^2\rho_0R_{0,eq}}{4M_{He}}\alpha_{eq}R_{0,eq}\frac{l^2+l+1}{2}.
\label{Eq:stiff}
\end{split}
$}
\end{equation}
It is clear that for large $l$, the stiffness coefficient $k_{lm}$ varies as $l^2$. Again, the summation of all the first-order terms in $\epsilon_{00}$ in Eq.~(\ref{Eq:E_S})-(\ref{Eq:E_m}) must vanish, and this can be used as a consistency check of the derivation.

\begin{table}[!tb]
\caption{Coefficients in the perturbation expansion for the molecular energy term $E_e$.}
\label{Table-1}
\begin{tabular}{p{0.7cm}|p{1.4cm}|p{1.4cm}|p{1.4cm}|p{1.4cm}|p{1.4cm}}
\hline\hline
T   & $k^{(a)}_0$ & $k^{(a)}_{00}$ & $k^{(a)}_{20}$ & $k^{(c)}_0$ & $k^{(c)}_2$\\
(K) & (cm$^{-1}$) & (cm$^{-1}$)    & (cm$^{-1}$)    & (cm$^{-1}$) & (cm$^{-1}$)\\
\hline
2.0 & -51.8       & 90.7           & 90.7           & -346.3      & -1005.9    \\
0.5 & -58.0       & 98.6           & 98.6           & -370.1      & -1074.4    \\
\hline\hline
\end{tabular}
\end{table}

When the He$^*_2$($a^{3}\Sigma^{+}_{u}$) bubble surface fluctuates, the liquid helium surrounding the bubble also moves. Following the method of Gross and Tung-Li \cite{Gross-1968-PR}, the kinetic energy \emph{KE} associated with the liquid motion can be derived as:
\begin{equation}
{KE}=\sum_{l,m}\frac{1}{2}M_{lm}|\dot{\epsilon}_{l,m}|^2,
\label{Eq:KE}
\end{equation}
where the effective mass of each mode $M_{lm}$ is given by:
\begin{equation}
M_{lm}=\frac{M_{He}\rho_0R^6_{0,eq}}{(l+1)R_{b,eq}}\bigg(1+\frac{4}{\alpha_{eq}R_{0,eq}}+\frac{6}{\alpha^2_{eq}R^2_{0,eq}}\bigg)^2.
\end{equation}
In this derivation, the liquid helium is treated as an incompressible ideal fluid \cite{Note-Incompressibility}.

\begin{figure}[!tb]
\includegraphics[scale=0.65]{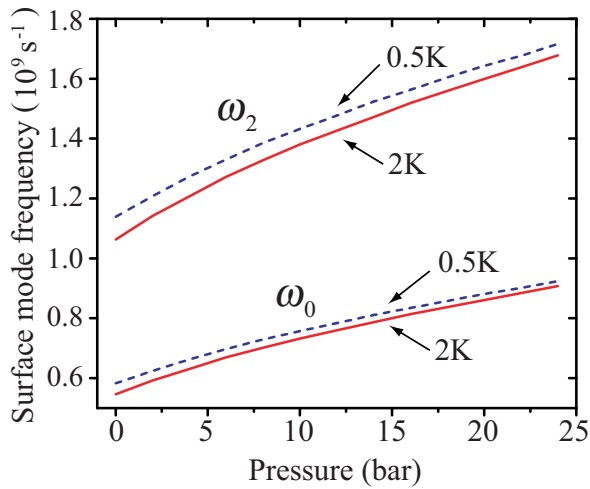}
\caption{(color online). Calculated frequencies of the $l$=0 and $l$=2 surface modes of the He$^*_2$($a^{3}\Sigma^{+}_{u}$) bubble as a function of helium pressure at 0.5K and 2 K.}\label{Fig3}
\end{figure}

Combining Eq. (\ref{Eq:dE}) and (\ref{Eq:KE}), one can construct a Lagrangian for the surface fluctuations:
\begin{equation}
\mathcal{L}=\sum_{l,m}\frac{1}{2}M_{lm}|\dot{\epsilon}_{l,m}|^2-\sum_{l,m}\frac{1}{2}k_{lm}|\epsilon_{lm}|^2.
\end{equation}
This Lagrangian represents a collection of harmonic oscillators in the parameter space $\{\epsilon_{lm}\}$, which are the eigenmodes of the bubble surface fluctuations. The frequency of each surface mode is given by $\omega_{l}$=$\sqrt{k_{lm}/M_{lm}}$. Note that these mode frequencies depend on $l$ but are independent of $m$. In Fig.~\ref{Fig3}, we show the calculated frequencies for the lowest two modes $l$=0 and $l$=2 as a function of pressure. The $l$=1 mode has zero frequency since it leads to a uniform translation of the entire bubble \cite{Maris-2004-JLTP}.

When calculating the mode frequencies, we adopt the pressure-dependent helium density as proposed by Maris and Edwards \cite{Maris-2002-JLTP}. In principle, a pressure-dependant surface tension should also be adopted. However, there is no reliable surface tension data at elevated pressures. On the other hand, it was found that the pressure dependance of the electron bubble absorption lines can be very well reproduced using pressure-independent surface tension values measured at saturated vapor pressures \cite{Golov-1995-ZPB}. Therefore, we use the experimentally measured surface tension at saturated vapor pressure in our calculations \cite{Note-Surf}. At a given temperature $T$, the mean amplitude of a surface mode $R_{0,eq}\langle\epsilon^2_{lm}\rangle^{1/2}$ can be evaluated via an ensemble average of both the zero-point and thermal fluctuations:
\begin{equation}
\langle\epsilon^2_{lm}\rangle=\frac{2}{k_{lm}}\bigg[\frac{1}{2}\hbar\omega_l+\frac{\hbar\omega_l}{e^{(\hbar\omega_l/k_BT)}-1}\bigg],
\end{equation}
where $k_B$ is the Boltzmann constant. At 2 K under saturated vapor pressure, we calculate that the mean amplitudes for the $l$=0 and $l$=2 modes are 0.46~{\AA} and 0.41 \AA, respectively, which justifies the fluctuation treatment of the bubble surface deformation. In the $T$=0 K limit, the mean amplitudes drop to 0.05 {\AA} and 0.06 {\AA} due to zero-point fluctuations.

\section{Optical transition of He$^*_2$ molecules}\label{Sec3}
In the perturbation framework, the cross-section $I(\omega)$ of the optical transition between two quantum states $|i\rangle$ and $|f\rangle$ is given by the Fermi's golden rule \cite{Shankar-book}:
\begin{equation}
I(\omega)\propto|\langle{f|\hat{z}|i}\rangle|^2\delta(E_f-E_i-\hbar\omega).
\end{equation}
This transition line is a delta function that peaks at the energy difference $E_f$--$E_i$ between the two states. Nonetheless, interactions between the quantum system and the environment can lead to the broadening of the line profile. A commonly adopted approach to account for this effect is the adiabatic line-broadening theory \cite{Anderson-1952-PR}. However, this approach is applicable in the static limit and does not appropriately account for the line broadening due to the zero-point and thermal fluctuations of the bubble shape. In what follows, we will adopt a different approach based on the surface eigenmodes of the He$^*_2$($a^{3}\Sigma^{+}_{u}$) bubbles. This method has been shown to well account for the observed broadening of the absorption lines for electron bubbles in liquid helium \cite{Maris-2004-JLTP, Guo-2007-JLTP}.

According to the Frank-Condon principle \cite{Franck-1926-TFS,Condon-1926-PR}, when a photon is absorbed the size and shape of the He$^*_2$($a^{3}\Sigma^{+}_{u}$) bubble should not change until \emph{after} the state of the electron wavefunction has changed. Therefore, for a deformed molecular bubble initially in the $a^{3}\Sigma^{+}_{u}$ state, the transition energy to the $c^{3}\Sigma^{+}_{g}$ state is:
\begin{equation}
\resizebox{0.88\hsize}{!}{$
\begin{split}
\Delta E_e&=\big(E^{(vac)}_{e,c}-E^{(vac)}_{e,a}\big)+\big(\int\big[U_c(\vec{r})-U_a(\vec{r})\big]\rho(\vec{r})d^3r\big)\\
&=\Delta E^{(vac)}_e+\Delta E_{shift}~,
\label{Eq:Tran-Energy}
\end{split}
$}
\end{equation}
where $\Delta E^{(vac)}_e$ is the $a$(0)${\rightarrow}$$c$(0) transition energy of a He$^*_2$ molecule in vacuum (with the corresponding excitation wavelength of 9183 \AA)\cite{Hill-1971-PRL}, and $\Delta E_{shift}$ is the shift in the transition energy due to the interaction between the molecule and the helium (which varies with the shape of the initial He$^*_2$($a^{3}\Sigma^{+}_{u}$) bubble). When the bubble shape fluctuates, an ensemble average of all possible initial shapes then leads to the broadening of the absorption line.

To describe the probability of a given bubble shape, we note that the probability density $P_l(\epsilon_{lm})$ of a surface mode with a displacement $\epsilon_{lm}$ is given by:
\begin{equation}
\resizebox{0.88\hsize}{!}{$
P_l(\epsilon_{lm})=\frac{\sum^{\infty}_{n=0}\psi^2_n(\epsilon_{lm})\exp\big[-(n+1/2)\hbar\omega_l/k_B T\big]}{\sum^{\infty}_{n=0}\exp\big[-(n+1/2)\hbar\omega_l/k_B T\big]},
$}
\label{Eq:Osc}
\end{equation}
where $\psi_n$'s are the eigenfunctions of the harmonic oscillator for $\epsilon_{lm}$. Eq.~(\ref{Eq:Osc}) can be simplified to \cite{Maris-2004-JLTP, Williams-1951-PR}:
\begin{equation}
\resizebox{0.88\hsize}{!}{$
P_l(\epsilon_{lm})=\sqrt{k_{lm}/2\pi k_BT^{(l)}_{eff}}\exp(-k_{lm}{\epsilon_{lm}}^2/2k_BT^{(l)}_{eff}),
$}
\label{Eq:Osc-Pro}
\end{equation}
where the mode effective temperature $T^{(l)}_{eff}$ is defined as $T^{(l)}_{eff}$=$\theta_l/\tanh(\theta_l/T)$, and $\theta_l$=$\hbar\omega_l/2k_B$. Therefore, through an ensemble average of all possible shape configurations $R_b(\theta,\phi)$, one can obtain the absorption line profile as:
\begin{equation}
\resizebox{0.88\hsize}{!}{$
I(\omega)=I_0\prod\limits_{l,m}\int d\epsilon_{lm}P_l(\epsilon_{lm})\delta\big(\Delta E_e\big(R_b(\theta,\phi)\big)-\hbar\omega\big),
\label{Eq:Spec}
$}
\end{equation}
where $I_0$ denotes the total absorption cross-section.

To the lowest order in $\epsilon_{lm}$, only the $l$=0 and $l$=2 surface modes can contribute to the shift of the $a$(0)${\rightarrow}$$c$(0) transition energy. The $l$=0 mode leads to an energy shift of the $a^{3}\Sigma^{+}_{u}$ state by $k^{(a)}_0\epsilon_{00}$ and an energy shift of the $c^{3}\Sigma^{+}_{g}$ state by $k^{(c)}_0\epsilon_{00}$. The $l$=2 mode only leads to an energy shift of the $c^{3}\Sigma^{+}_{g}$ state energy by $k^{(c)}_2\epsilon_{20}$. As a result, the absorption line profile can be simplified to:
\begin{equation}
\begin{split}
&I(\omega)=I_0\int d\epsilon_{00}P_0(\epsilon_{00})\int d\epsilon_{20}P_2(\epsilon_{20})\\
&~~\times\delta\big(\Delta E_{e,eq}+(k^{(c)}_0-k^{(a)}_0)\epsilon_{00}+k^{(c)}_2\epsilon_{20}-\hbar\omega\big),
\label{Eq:Spec-sim}
\end{split}
\end{equation}
where $\Delta{E_{e,eq}}$ denotes the transition energy of the He$^*_2$($a^{3}\Sigma^{+}_{u}$) bubble at a given temperature and pressure without any shape deformation. Using Eq.~(\ref{Eq:Osc-Pro}) for $P_0(\epsilon_{00})$ and $P_2(\epsilon_{20})$, we can integrate the above equation:
\begin{equation}
\resizebox{0.88\hsize}{!}{$
I(\omega)=I_0\exp\left(-\frac{(\Delta{E_{e,eq}}-\hbar\omega)^2}{\frac{(k^{(c)}_0-k^{(a)}_0)^2}{k_{00}}2k_BT^{(0)}_{eff}+\frac{(k^{(c)}_2)^2}{k_{20}}2k_BT^{(2)}_{eff}}\right).
$}
\end{equation}
Therefore, the absorption line to the lowest order in $\epsilon_{lm}$ is a Gaussian, and its broadening essentially comes from the $l$=0 and $l$=2 surface modes. We would like to point out that in Ref.~\cite{Kanorsky-1994-PRB}, the authors only considered the $l$=0 mode when calculating the absorption line profile for barium atomic bubbles in liquid helium and therefore obtained results that disagreed with observations.

\begin{figure}[!tb]
\includegraphics[scale=0.62]{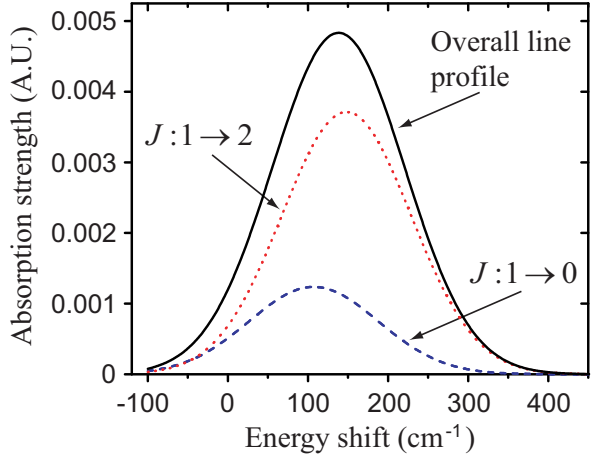}
\caption{$a$(0)${\rightarrow}$$c$(0) absorption line at 2 K under zero applied pressure. The blue dashed curve and the red dotted curve are the absorption lines due to the $P$ branch transition $J$:1${\rightarrow}$0 and $R$ branch transition $J$:1${\rightarrow}$2. The black solid curve is the overall absorption line with the total area normalized to unity.}
\label{Fig4}
\end{figure}

In the above analysis, the motion of the molecular ion core $^+$He$^*_2$ inside the bubble is not considered. There are two major effects of this motion on the $a$(0)${\rightarrow}$$c$(0) absorption transition. The first is the oscillations of the ion core around its equilibrium position, which can contribute to the broadening of the absorption line. However, the authors of Ref.~\cite{Eloranta-2002-JCP} showed that this additional broadening is only about 10 cm$^{-1}$, which is far smaller than the broadening due to the bubble shape fluctuations. Another effect of the ion core motion is that in the He$^*_2$($a^{3}\Sigma^{+}_{u}$) state, due to the nearly spherical equilibrium shape of the bubble, the ion core can rotate in the bubble like a free rotor. The associated rotational energy $E_{rot}$ is approximately given by $E_{rot}$=$B{\cdot}J(J+1)$, where $B$=7.6 cm$^{-1}$ is the rotational constant \cite{Herzberg-book,Ginter-1965-JCP} and the integer $J$ denotes the angular quantum number of the rotational level. The He$^*_2$ molecule in different rotational levels of the $a^{3}\Sigma^{+}_{u}$ state can be excited to the corresponding rotational levels of the $c^{3}\Sigma^{+}_{g}$ state according to the selection rule $\Delta J$=+1 ($R$ branch) and $\Delta J$=-1 ($P$ branch). Due to the internal symmetry, $J$ must be an odd number for the $a^{3}\Sigma^{+}_{u}$ state but an even number for the $c^{3}\Sigma^{+}_{g}$ state. At the temperatures of interests, the fraction of the $a^{3}\Sigma^{+}_{u}$ molecules occupying the $J$=1 level is far larger than that in the other levels \cite{Hill-1971-PRL}. As a result, practically we only need to consider the $a$(0)${\rightarrow}$$c$(0) transition with $J$:1${\rightarrow}$0 and $J$:1${\rightarrow}$2. The additional energy shift associated with the change in the rotational levels is $-2B$ for $J$:1${\rightarrow}$0 and $4B$ for $J$:1${\rightarrow}$2. The overall $a$(0)${\rightarrow}$$c$(0) absorption line profile is then the summation of the two transition lines with their statistical weight (i.e., 1:3). \cite{Eloranta-2002-JCP} As an example, we show the calculated $a$(0)${\rightarrow}$$c$(0) absorption line at 2 K under zero applied pressure in Fig.~\ref{Fig4}.

\begin{figure}[!tb]
\includegraphics[scale=0.6]{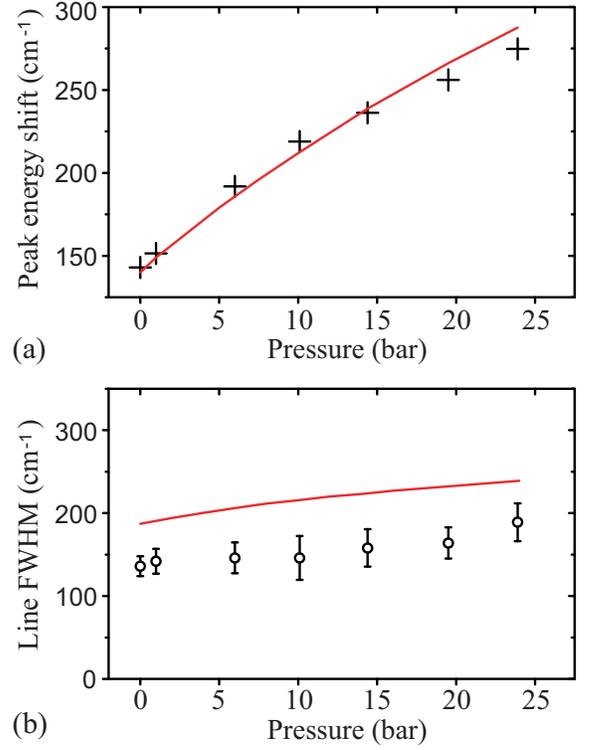}
\caption{(a) The peak location and (b) the FWHM of the overall $a$(0)${\rightarrow}$$c$(0) absorption line as a function of pressure at 2 K. The red curves represent the calculated results. The crosses and circles are experimental data of Eltsov \emph{et al.} \cite{Eltsov-1998-JLTP}, taken in the temperature range of 1.76 K to 2.05 K.} \label{Fig5}
\end{figure}
\begin{figure}[htb]
\includegraphics[scale=0.62]{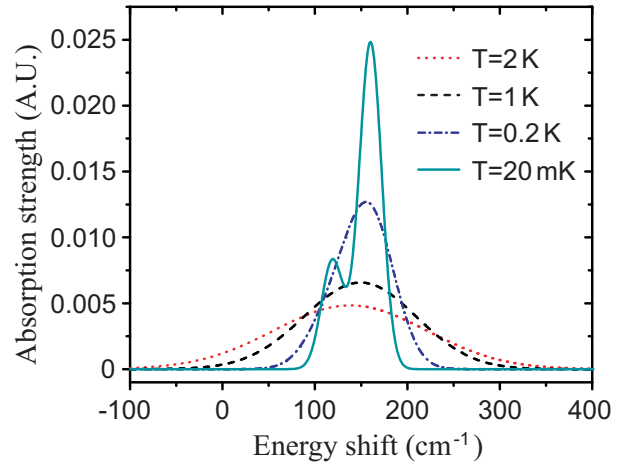}
\caption{Calculated overall $a$(0)${\rightarrow}$$c$(0) absorption line profiles at various temperatures. The total area below each curve is normalized to unity} \label{Fig6}
\end{figure}
In order to validate our calculation, in Fig.~\ref{Fig5} we show the obtained peak location and the FWHM of the overall $a$(0)${\rightarrow}$$c$(0) absorption line as a function of the applied pressure at 2 K, together with the experimental data of Eltsov \emph{et al.} \cite{Eltsov-1998-JLTP}. The calculated peak location shows an excellent agreement with the experimental data. The obtained FWHM of the lines appear to be consistently larger than the measured ones. Nevertheless, considering the quality of the experimental absorption lines and the relatively large line width as seen in Fig.~\ref{Fig4}, we regard the agreement as reasonable.

In Fig.~\ref{Fig6}, we show the calculated line profiles at various temperatures under saturated vapor pressures. It is interesting to see that as the temperature decreases, the width of the absorption line also decreases. This can be understood as due to the reduced thermal fluctuations of the He$^*_2$($a^{3}\Sigma^{+}_{u}$) bubble shape at low temperatures. Nonetheless, even at zero temperature, the $a$(0)${\rightarrow}$$c$(0) absorption line should still have a finite width due to the zero-point fluctuations of the $l$=0 and $l$=2 modes. At the lowest temperature shown in Fig.~\ref{Fig6}, due to the much reduced line width, the two peaks due to the rotational levels can be clearly resolved. From 2 K to 20 mK, the maximum strength of the absorption line is enhanced by a factor of about five. Besides, there is a blueshift of the peak transition energy by about 22 cm$^{-1}$. If we take 910 nm as the peak $a$(0)${\rightarrow}$$c$(0) excitation wavelength at 2 K, this energy shift suggests that the peak excitation wavelength at 20 mK will be about 908 nm.

\section{Diffusion of He$^*_2$ molecules in low-temperature He II}\label{Sec4}
Neutral He$^*_2$ molecules move diffusively in He II due to collisions with thermal quasiparticles (phonons and rotons). Below about 0.6 K, this diffusion is essentially controlled by He$^*_2$-phonon scattering. At sufficiently low temperatures when their mean free path through the thermal phonons becomes comparable or greater than the size of the helium container, the motion of the He$^*_2$ molecules can become ballistic~\cite{Zmeev-2013-JLTP}. Knowing the diffusion coefficient $D$ of the He$^*_2$ molecules in low-temperature He II is important for the design of future vortex-line imaging experiments. A rough estimate of $D$ was provided by McKinsey \emph{et al.} \cite{McKinsey-2005-PRL}, but their suggested value at 0.2 K appears to be three orders of magnitude greater than the one extracted from Zmeev \emph{et al.}'s measurement \cite{Zmeev-2013-JLTP}. Following Baym \emph{et al.} \cite{Baym-1969-PRL}, hereby we provide a more realistic evaluation of $D$ by considering the momentum-transfer in He$^*_2$-phonon scattering.

The diffusion coefficient $D$ is related to the mobility $\mu$ of a particle through the Einstein-Smoluchowski relation $D$=$\mu k_BT$, where $\mu$=$v_d/F$ is defined as the ratio of the particle's terminal velocity $v_d$ to an applied force $F$. According to Baym \emph{et al.} \cite{Baym-1969-PRL}, when phonon scattering dominates the energy dissipation, the particle mobility in He II is given by:
\begin{equation}
{\mu}^{-1}=-\frac{\hbar}{6\pi^2}\int^{\infty}_0dk k^4\frac{\partial n(k,T)}{\partial k}\sigma_T(k)
\label{Eq:Mobility}
\end{equation}
where $n(k,T)=[\exp(\hbar v_ck/k_BT)-1]^{-1}$ is the equilibrium phonon distribution function, $v_c$ is the sound velocity in He II, and $\sigma_T(k)$ denotes the momentum-transfer cross section for incident phonons at a wave number $k$. $\sigma_T(k)$ can be evaluated as:
\begin{equation}
\sigma_T(k)=\int d\Omega(1-\cos\theta)\sigma(k,\theta),
\end{equation}
where the differential cross section $\sigma(k,\theta)$ for phonon scattering through an angle $\theta$ is given by \cite{Baym-1969-PRL}:
\begin{equation}
\sigma(k,\theta)=k^{-2}|\sum^\infty_{l=0}(2l+1)P_l(\cos\theta)f_l(k)|^2.
\label{Eq:micro-cross}
\end{equation}
Here $f_l(k)$ denotes the amplitude of the $l^{th}$ outgoing spherical wave for an incident planar wave at $k$. For a He$^*_2$($a^{3}\Sigma^{+}_{u}$) bubble, $f_l(k)$ is given by~\cite{Celli-1968-PR}:
\begin{equation}
f_l(k)=i\frac{j'_l(kR_{b,eq})+\gamma_lkR_{b,eq}j_l(kR_{b,eq})}{h'_l(kR_{b,eq})+\gamma_lkR_{b,eq}h_l(kR_{b,eq})},
\end{equation}
where $j_l$ and $h_l$ are the spherical Bessel and Hankel functions, and the prime denotes the derivative. The coefficient $\gamma_l$ is related to the surface mode stiffness $k_l$ of the He$^*_2$($a^{3}\Sigma^{+}_{u}$) bubble as:
\begin{equation}
\gamma_l=\frac{M_{He}\rho_0v^2_cR^6_{0,eq}}{k_lR^3_{b,eq}}\big(1+\frac{4}{\alpha_{eq}R_{0,eq}}+\frac{6}{\alpha^2_{eq}R^2_{0,eq}}\big)^2.
\end{equation}
Using Eq.~(\ref{Eq:stiff}), we find that $\gamma_0$=$0.38$ and $\gamma_2$=$0.3$, nearly independent of temperature at $T$$<$0.5 K under saturated vapor pressure. These values are about an order of magnitude smaller than those for electron bubbles in He II. This is due to the smaller size of the He$^*_2$($a^{3}\Sigma^{+}_{u}$) bubble and its larger surface mode elastic constant $k_l$. For the $l$=1 contribution in Eq.~(\ref{Eq:micro-cross}), one can take $f_1(k)$=$ij_1(kR_{b,eq})/h_1(kR_{b,eq})$ since $\gamma_1$ diverges due to the vanishing mode stiffness \cite{Baym-1969-PRL}, $k_1$=0.

\begin{figure}[!tb]
\includegraphics[scale=0.6]{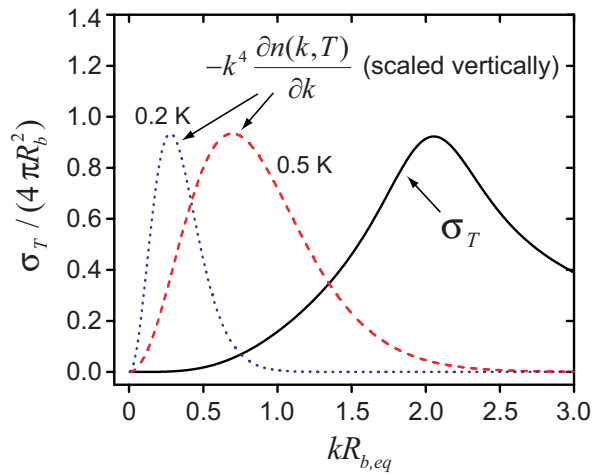}
\caption{The momentum-transfer cross section $\sigma_T(k)$ calculated using the $l$=0,1,2 modes. The dotted and dashed curves represent the thermal factor $-k^4\partial n/\partial k$, which are scaled vertically for visibility.}\label{Fig7}
\end{figure}

In Fig.~\ref{Fig7}, we show the profile of $\sigma_T(k)$ together with the thermal factor $-k^4\partial n(k,T)/\partial k$. The $\sigma_T(k)$ is calculated based on the contributions from the $l$=0, 1, 2 wave scattering. Adding more modes only changes the right tail part of the profile. Contrary to electron bubbles, the profile of $\sigma_T(k)$ for He$^*_2$($a^{3}\Sigma^{+}_{u}$) bubbles does not exhibit any sharp peaks due to resonant phonon scattering. Furthermore, the dominant contribution to the left tail of $\sigma_T(k)$ comes from the $l$=1 scattering. Evidently, the thermal factor peaks at $kR_{b,eq}<$1 and shifts towards the origin as $T$ decreases. Therefore, when we compute the mobility $\mu$ using Eq.~(\ref{Eq:Mobility}), the major contribution at low temperatures comes from the $l$=1 wave scattering.

\begin{figure}[!tb]
\includegraphics[scale=0.62]{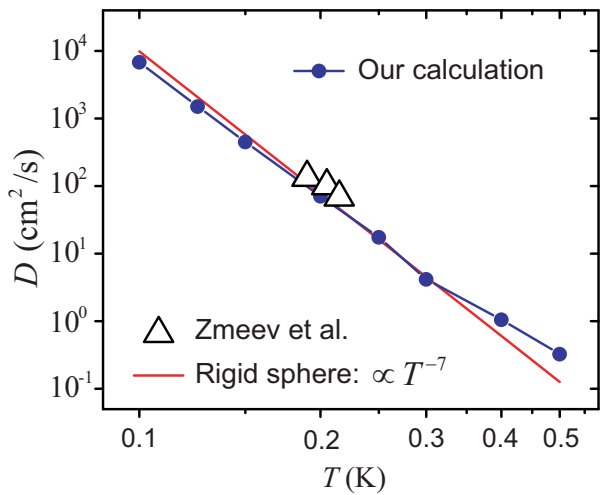}
\caption{Calculated diffusion coefficient $D$ of He$^*_2$($a^{3}\Sigma^{+}_{u}$) molecules in He II. The triangles represent the data extracted from Zmeev \emph{et al.}'s experiment \cite{Zmeev-2013-JLTP}. The red solid line is the diffusion curve calculated using the rigid sphere model as discussed in the text.}\label{Fig8}
\end{figure}

In Fig.~\ref{Fig8}, we show the calculated He$^*_2$ diffusion coefficient $D$ based on the obtained mobility result. The experimental data around 0.2 K, extracted from the He$^*_2$ diffusion time measurement by Zmeev \emph{et al.}~\cite{Zmeev-2013-JLTP}, are also included. It is clear that our calculated result agrees quite well with the measurement, which proves the reliability of our calculations. For comparison purpose, we have also included in Fig.~\ref{Fig8} the calculated $D$ coefficient assuming the He$^*_2$ bubble as a rigid sphere, i.e., $\gamma_l$=0 for all modes with $l\neq$1. At $T$ less than about 0.3 K, the solid sphere model appears to agree well with our earlier calculation based on the $l$=0, 1, 2 scattering modes, which suggests that in this temperature regime the He$^*_2$ bubble can be reasonably treated as a rigid sphere due to its large shape deformation stiffness. In the low $T$ limit, the value of $D$ for the rigid sphere model is always greater by a factor of about 1.6. Indeed, in the low $T$ limit where the thermal factor peaks at $kR_b\ll$1, an analytic expression for $\mu$ can be derived \cite{Baym-1969-PRL}:
\begin{equation}
\mu=\frac{1}{(2/9+\gamma_0^2)\rho_nv_c4\pi R^2_{b,eq}}\bigg(\frac{\hbar v_c}{2\pi R_{b,eq}k_BT}\bigg)^4,
\end{equation}
where $\rho_n$=$2\pi^2{k_B}^4T^4/45\hbar^3v^5_c$ is the mass density of the normal fluid in He II due to the phonon contribution~\cite{Wilks-1967-book}. Hence, the resulted diffusion coefficient is:
\begin{equation}
D=\frac{45\hbar^7v_c^8}{(2/9+\gamma_0^2)128\pi^7R^6_{b,eq}}(k_BT)^{-7}.
\end{equation}
Therefore, if the contribution of $\gamma_0$ is neglected, the value of $D$ would increase by a factor of 1.6.

\section{Summary}\label{Sec5}
We have derived the surface fluctuation eigenmodes for the He$^*_2$($a^{3}\Sigma^{+}_{u}$) excimer molecules in He II, using a bubble model that incorporates the He$^*_2$-He interaction potentials obtained in density functional analysis. These eigenmodes are then utilized in the evaluation of the line profile of the $a^{3}\Sigma^+_{u}{\rightarrow}c^{3}\Sigma^+_{g}$ absorption transition that controls the fluorescence imaging of the He$^*_2$ molecules. We find that as the temperature drops from 2 K to 20 mK, the peak absorption strength is enhanced by a factor of about five and the optimum excitation wavelength is blueshifted by about 2 nm. We have also calculated the diffusion coefficient of the He$^*_2$ molecules in He II at temperatures below 0.5 K by considering the momentum transfer in He$^*_2$-phonon scattering. The good agreement between our result and the experimental data obtained at about 0.2 K proves the reliability of our calculation. Our analysis suggests that due to the large shape deformation stiffness, the He$^*_2$($a^{3}\Sigma^{+}_{u}$) bubble can be reasonably treated as a rigid sphere below about 0.3 K when they scatter off thermal phonons. These results will provide a useful guidance in the design of future vortex-line imaging experiment in He II using He$^*_2$($a^{3}\Sigma^{+}_{u}$) molecules as tracers.

\begin{acknowledgments}
W. G. acknowledges the support by the National Science Foundation (NSF) under Grant No. DMR-1807291 and the support by U.S. Department of Energy under Grant No. DE-SC0020113. The work was partly done at the National High Magnetic Field Laboratory which is supported through the NSF Cooperative Agreement No. DMR-1644779 and the state of Florida. W. G. would also like to thank the Department of Physics and Astronomy at the University of Manchester for hosting his sabbatical visit. Both A. I. G. and W. G. acknowledge the support provided by the Engineering and Physical Sciences Research Council (EPSRC) in United Kingdom through the Grant No. EP/P025625/1.
\end{acknowledgments}

\end{document}